\documentclass[twocolumn,showpacs,preprintnumbers,amsmath,amssymb]{revtex4}

\usepackage{graphicx}
\usepackage{dcolumn}
\usepackage{bm}
\begin{document}

\title{Anisotropic electronic transport and Rashba effect of the
two-dimensional electron system in (110) SrTiO$_3$-based heterostructures}
\author{K. Wolff, R. Eder, R. Sch\"afer, R. Schneider, and D. Fuchs}
\affiliation{Karlsruher Institut f\"ur Technologie, Institut f\"ur
Festk\"orperphysik, 76021 Karlsruhe, Germany }
\date{\today}
\begin{abstract}
The two-dimensional electron system in $(110)$ Al$_2$O$_{3-\delta}$/SrTiO$_3$
heterostructures displays anisotropic electronic transport. Largest and lowest
conductivity and electron mobility $\mu$ are observed along the $[001]$ and
$[1\bar{1}0]$ direction, respectively. The anisotropy of the sheet resistance 
and $\mu$ likewise leads to a distinct anisotropic normal magnetotransport 
($M\hspace{-0.5mm}R$) for $T < 30$K. However, at temperatures
$T<5$K $M\hspace{-0.5mm}R$ and magnetic field $B< 2$T $M\hspace{-0.5mm}R$
is dominated by weak antilocalization.
Despite the rather strong anisotropy of the
Fermi surfaces, the in-plane anisotropic magnetoresistance
($AM\hspace{-0.5mm}R$) displays two-fold non-crystalline anisotropy. However, 
the $AM\hspace{-0.5mm}R$-amplitude is found to be anisotropic with respect to 
the current direction, leading to a 60\% larger $AM\hspace{-0.5mm}R$ amplitude 
for current $I$ along the $[001]$ direction compared to $I$ parallel to
$[1\bar{1}0]$. 
Tight binding calulations evidence an anisotropic Rashba-induced
band splitting with dominant linear $k$-dependence. In combination with
semiclassical Boltzmann theory  the non-crystalline $AM\hspace{-0.5mm}R$ 
is well described, despite the anisotropic Fermi surface.
\end{abstract}
\maketitle
\section{Introduction}
The two-dimensional electron system (2DES) formed at the interface of
the band insulators LaAlO$_3$ (LAO) and SrTiO$_3$ (STO) displays many
intriguing features such as superconductivity, spin-orbit interaction (SOI)
and multiple quantum criticality \cite{1,2,3}, and has thus made LAO/STO a
prototypical system for studying low-dimensional strongly correlated electron
systems. Magnetic properties are reported alike \cite{4}, however, believed to
arise rather from extrinsic sources like oxygen vacancies and strain. \\
In (001) oriented LAO/STO the sheet carrier concentration $n_s$ can be tuned by
electric field gating through a Lifshitz transition \cite{5} occurring at a
critical sheet carrier concentration $n_c \approx 1.7\times10^{13} $ cm$^{-2}$,
where itinerant electrons change from populating only Ti derived 3d
$t_{2g}$ orbitals with $d_{xy}$ symmetry to occupying also the
$d_{xz}$, $d_{yz}$ orbitals. These bands result in a highly elliptical
Fermi
surface oriented along crystalline directions and may give reason for the
observation of crystalline anisotropic electronic properties. In addition,
localized magnetic moments, pinned to specific $d_{xy}$ orbitals may lead to
crystalline
 anisotropy as well and may complicate anisotropic electronic transport. The
coexistence of localized charge carriers close to the interface and itinerant
$d$ electrons may lead to fascinating phenomena such as non-isotropic
magnetotransport or magnetic exchange. However, it is not clear whether
interaction between these localized magnetic moments and mobile charge
carriers really happens. \\
The SOI in (001) LAO/STO results in a non-crystalline two-fold anisotropic
in-plane magnetoresistance ($AM\hspace{-0.5mm}R$) \cite{6}. Interestingly,
for $n_s > n_c$ sometimes
a more complex $AM\hspace{-0.5mm}R$ with a four-fold crystalline anisotropy
is reported which
is discussed in terms of a tunable coupling between itinerant 
electrons and electrons localized in
$d_{xy}$ orbitals at Ti vacancies\cite{7}. However, the appearance of a
crystalline $AM\hspace{-0.5mm}R$ with increasing $n_s$ is not always evident
and rises question
about its microscopic origin. More recently, a giant crystalline
$AM\hspace{-0.5mm}R$ of up
to 100\% was reported in (110) oriented LAO/STO \cite{8}. Here, $n_s$ was
about
$2\times10^{13}$ cm$^{-2}$. However, the $AM\hspace{-0.5mm}R$ was
supposed to be related to strong
anisotropic spin-orbit field and the anisotropic band structure of (110)
LAO/STO.\\
With respect to both, namely fundamental aspects such as the possible
simultaneous
appearance of magnetism and superconductivity and applications in the field
of spintronics, a more fundamental knowledge about the origin of anisotropic
magnetotransport is highly desired. Measurements of the $AM\hspace{-0.5mm}R$
in a rotating
in-plane magnetic field are well suited to probe crystalline anisotropy and
symmetry of a 2DES and are a promising tool to elucidate magnetic properties
because of its high sensitivity towards spin-texture and spin-orbit
interaction \cite{9}. \\
In order to investigate the microscopic origin of the anisotropic electronic
properties of the 2DES of STO-based heterostructures we studied in detail
the electronic transport of the 2DES formed at the interface of
spinel-type Al$_2$O$_{3-\delta}$ and (110) oriented STO (AO/STO). The presence
of oxygen vacancies \cite{10} promoting localized $d_{xy}$ electrons in
combination
with the anisotropic band structure of (110) STO surface \cite{11} makes (110)
AO/STO very suitable for these experiments. The heterostructures were produced
by standard pulsed laser deposition, and characterization of the electronic
transport was done by sheet resistance measurements. Band structure
calculations
were carried out using linear combination of atomic orbitals (LCAO)
approximation to model band structure and Fermi surface properties of
(110) AO/STO. $AM\hspace{-0.5mm}R$ was deduced using
semi-classical Boltzmann theory. Surprisingly, despite the anisotropy of the
electronic band structure and SOI, and the presumably
large content of oxygen vacancies as compared to LAO/STO, we did not observe
indications for a crystalline $AM\hspace{-0.5mm}R$ in (110) AO/STO. The 
$AM\hspace{-0.5mm}R$ displays two-fold non-crystalline anisotropic behavior.
Contributions to the electronic transport from the different Fermi surface
sheets as well as the anisotropy of the Fermi surfaces itself are sensitively
affected by $n_s$.
\section{Experimental}
\begin{figure}
\includegraphics[width=0.9\columnwidth]{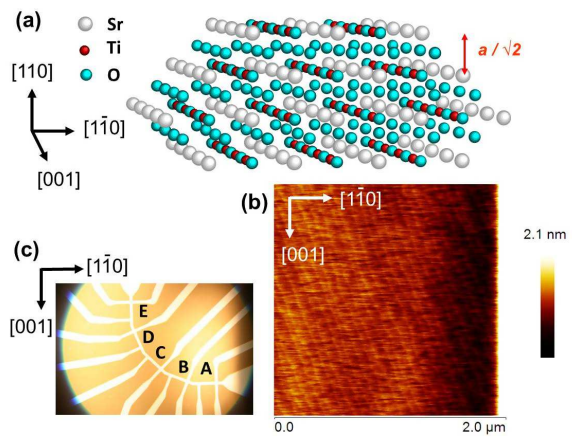}
\caption{\label{fig1}
(a) Schematic of the crystal structure of STO. In case of $(110)$
orientation, the surface can be terminated with a SrTiO or oxygen layer.
The spacing of the cation layers is $a/\sqrt{2}$, where $a = 3.905$\AA$\;$ is
the cubic lattice parameter
of STO. Crystallographic directions and atom labels are indicated. (b) Surface
topography before Al$_2$O$_{3-\delta}$ deposition characterized by atomic force
microscopy. The image was taken on microbridge A. (c) Optical micrograph of a
patterned sample. Sharp contrast between AO/CeO$_2$ (dark) and AO/STO (bright)
enables identifying  microbridges labeled alphabetically from A$-$E. }
\end{figure}

Sample preparation has been carried out by depositing Al$_2$O$_{3-\delta}$ films
onto $(110)$ oriented STO substrates
with  a thickness of about 15 nm at a substrate temperature of
$T_s = 250^{\circ}C$  by pulsed laser deposition \cite{12}. In order to achieve
an atomically flat, single-type terminated substrate surface, the substrates
are annealed at $T = 950^{\circ}C$ for 5h in flowing oxygen.
The $(110)$ STO
surface can be terminated by a SrTiO or an oxygen layer, see
Fig. \ref{fig1} (a), where
the cation composition at the interface should be always the same in case of
single-type termination. Annealing results in a stepped surface topography
with a step height of about $2.7$\AA $\;$and a step width of 80 nm, see Fig.
\ref{fig1} (b).
Oxygen partial pressure during Al$_2$O$_{3-\delta}$ deposition and
cool-down process was
$p(O_2) = 10^{-6}$ mbar. Prior to the deposition, microbridges with
a
length of $100 \mu$m and a width of $20 \mu$m in Hall bar geometry have been
patterned along specific crystallographic directions using a CeO$_2$ hard mask
technique \cite{13}, see Fig. \ref{fig1} (c).
The microbridges are labeled from A to E, with angle
$\varphi = 0^{\circ}, 22.5^{\circ}, 45^{\circ}, 67.5^{\circ}$, and $90^{\circ}$
towards the $[1\bar{1}0]$ direction, i.e.,
A and E parallel to $[1\bar{1}0]$ and $[001]$ direction, respectively.\\
The sheet resistance $R_s$ was measured using a physical property measurement
system (PPMS) from Quantum Design in the temperature and magnetic field ranges
$2$ K $\le T \le 300$K and $0 \le B \le 14$T. To avoid charge carrier
activation by
light \cite{14,15}, alternating current measurements ($I_{ac} = 3 \mu$A) were
started
not before 12 hours after loading the samples to the PPMS.  The
magnetoresistance, $M\hspace{-0.5mm}R = [R_s(B) - R_s(0)]/R_s(0)$, and the
$AM\hspace{-0.5mm}R =[R_s(B_{ip},\phi) - R_s(B_{ip},0)]/R_s(B_{ip},0)$, 
have been measured with magnetic
field  normal ($B$) and parallel ($B_{ip}$) to the interface, respectively. For
measuring
$AM\hspace{-0.5mm}R$ with rotating in-plane magnetic field $B_{ip}(\phi)$, a 
sample rotator was used.
The angle $\phi$ between $B_{ip}$ and $[001]$-direction was varied from
$0^{\circ} - 360^{\circ}$.
Special care has been taken to minimize sample wobbling in the apparatus. 
Residual
tilts ($1^{\circ} - 2^{\circ}$) of the surface normal with respect to the rotation
axis which
produces a perpendicular field component oscillating in sync with $\phi$ could
be identified by comparison of $R_s(B_{ip},\phi)$ for different microbridges and
could therefore be corrected properly.
\section{Results and Discussion}
\subsection{Temperature dependence of the anisotropic electronic transport}
\begin{figure}
\includegraphics[width=\columnwidth]{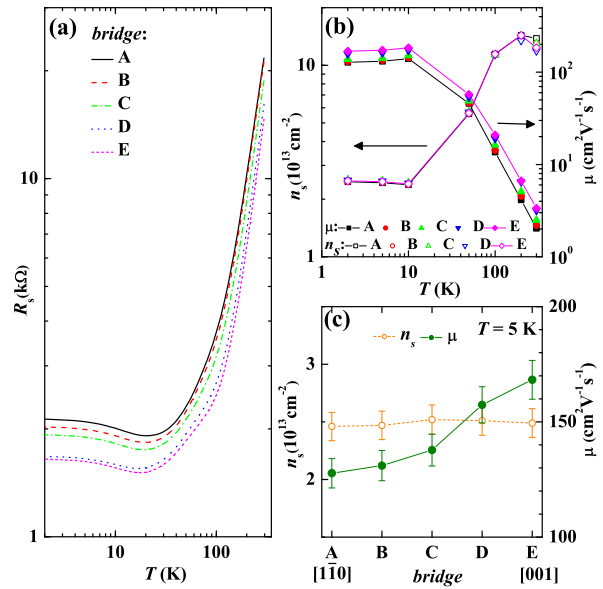}
\caption{
(a) Sheet resistance $R_s$ versus $T$ for microbridges A $-$ E (from top
to bottom) with an angle
$\varphi = 0^{\circ}, 22.5^{\circ}, 45^{\circ}, 67.5^{\circ}$, and $90^{\circ}$
towards the $[1\bar{1}0]$ direction, i.e., $A \parallel [1\bar{1}0]$ and
$E \parallel [001]$. (b) Sheet carrier density $n_s$, left scale, and Hall
mobility $\mu$, right scale, versus $T$ for A $-$ E. (c) $n_s$, left scale,
and $\mu$, right scale for bridge A$-$E at $T = 5$K.}
\label{fig2}
\end{figure}

The anisotropic electronic band structure of the 2DES found in $(110)$ oriented
LAO/STO heterostructures \cite{8} and at the reconstructed surface of $(110)$
oriented STO \cite{11} obviously lead to anisotropic electronic transport
\cite{16,17}. The lowest electronic subbands along the $[1\bar{1}0]$
direction (along $\Gamma-M$) display much weaker dispersion and smaller
band-width compared to
the $[001]$ direction (along $\Gamma-Z$) which typically results in larger
resistance
for current $I$ direction along the $[1\bar{1}0]$ direction \cite{17,8}. The
electronic transport in $(110)$ AO/STO displays distinct anisotropy as well.
The $T$-dependence of the sheet resistance $R_s$ along different
crystallographic
 directions is shown in Fig. \ref{fig2} (a). For all the microbridges, 
$R_s$ decreases
with deceasing $T$ nearly $\propto T^2$ down to about 100 K and shows a shallow
minimum
around 20 K. The $T$-dependence is very similar to that observed in $(001)$
AO/STO
and is likely explained by strong renormalization due to electron-phonon
interaction and impurity scattering \cite{18,19}. The resistivity ratio
between 300 K and 10 K amounts to about 20, which is nearly the same as
that of $(001)$ AO/STO. $R_s$ steadily decreases from A
($I \parallel [1\bar{1}0]$) to E ($I \parallel [001]$) with
increasing $\varphi$ at constant $T$
throughout the complete $T$-range. Obviously, anisotropic transport is not only
restricted to low temperatures $T < 10 K$, where usually impurity
scattering dominates $R_s$. Moreover, the anisotropy between A and E,
$[R_s(A) - R_s(E)]/R_s(E)$ is largest at $T = 300$K amounting to 47\% and
decreases with decreasing $T$ to 29\% at $T = 5$K. This rather small 
$T$-dependence
indicates that the intrinsic anisotropic electronic band structure is very
likely the dominant source for the anisotropic transport. In contrast,
anisotropic transport in $(001)$ AO/STO is extrinsic in nature and 
is found only at low $T$ where it is
caused mainly by anisotropic impurity scattering
due to an inhomogeneous distribution
of $<110> $lattice dislocations \cite{20}. With respect to these results,
anisotropy in $(110)$ AO/STO may be diminished at low $T$ and intrinsic
anisotropy would be even larger. \\
In order to extract sheet carrier density $n_s$ and mobility $\mu$,
Hall measurements have been carried out in a magnetic field 
$-14$T$\le B \le 14$T
applied normal to the interface for $2$K$\le T \le 300$K. For $T < 30$K, the
Hall resistance $R_{xy}$ becomes slightly nonlinear, indicating multi-type
carrier transport. However, $n_s$ which we determined from the asymptotic
value of $R_{xy}$ at high fields, i.e., the total $n_s$, usually deviates by
less than 10\% from $n_s$ extracted from $R_{xy}$ in the limit of $B = 0$.
In Fig. \ref{fig2} (b) the total $n_s$ and the Hall mobility, calculated by
$\mu = (R_s(B = 0)\times n_s\times e)^{-1}$, where $e$ is the elementary charge,
are shown as functions of $T$. $n_s$ decreases with decreasing $T$ from about
$1.3\times 10^{14} $ cm$^{-2}$ at $T = 300$K to $2.5\times 10^{13}$ cm$^{-2}$ at
$T = 5$K and is well comparable to that of $(110)$ LAO/STO \cite{10}. \\
In contrast to the $T$-dependence of $n_s$, $\mu$ increases from about
$2.5$ cm$^2/($Vs$)$ with decreasing $T$ to $150$ cm$^2/($Vs$)$. The
$T$-dependence of
$n_s$ and $\mu$ is well comparable to that observed in 2DES of $(001)$ STO
based heterostructures \cite{21,22,20}. As expected from $R_s$ the maximum
anisotropy of $n_s$ and $\mu$ is observed at $T = 300$K amounting to 16\% and
65\%, respectively, and decreases to 2\% and 34\% at $T = 5$K. Therefore, the
anisotropy of $R_s$ at low $T$ is mainly caused by the anisotropy of $\mu$,
whereas $n_s$
for the different microbridges A$-$E are roughly the same. The superior role
of $\mu$ with respect to electronic anisotropy is demonstrated in
Fig. \ref{fig2} (c)
where  $n_s$ and $\mu$ are plotted for A$-$E at $T$ = 5K. $n_s$ differs only
little
for the different microbridges. In contrast, $\mu$ steadily increases from A
to E with increasing $\varphi$ and shows the highest mobility for bridge E,
i.e., along the $[001]$ direction. The results are reasonable with respect
to the anisotropic band structure and Fermi surface of $(110)$ AO/STO, which
will be discussed in more detail in III. C. \\
\subsection{Magnetotransport}
\begin{figure}
\includegraphics[width=0.6\columnwidth]{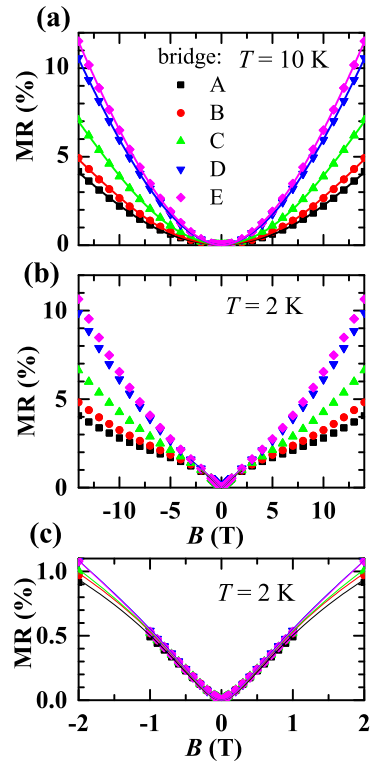}
\caption{\label{fig3}
Magnetoresistance $M\hspace{-0.5mm}R$ for the different microbridges 
A$-$E at $T = 10 $K
 (a) and $2$K (b) versus magnetic field $B$ applied perpendicular to the
interface. (c) $M\hspace{-0.5mm}R$ at $T = 2$K in the low-field range for 
$B < B^*$.
Fits to the data with respect to the Kohler form and MF-expression
are shown by solid lines, see text.}
\end{figure}

\begin{figure}
\includegraphics[width=\columnwidth]{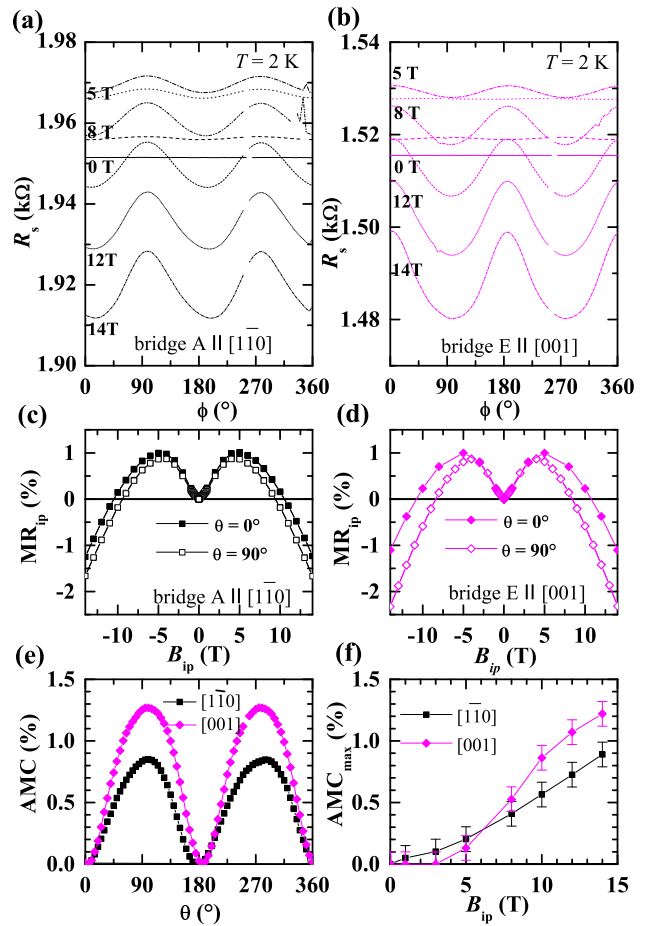}
\caption{\label{fig4}
$AM\hspace{-0.5mm}R$ of $(110)$ AO/STO at 2 K. $R_s$ versus in-plane angle
$\phi$ for
different strengths of $B_{ip}$ $(0, 1, 3, 5, 8, 10, 12$, and $14$T$)$ for (a)
bridge A, i.e., $I$ parallel to the $[1\bar{1}0]$ direction, and (b) bridge E,
$I$ being parallel to the $[001]$ direction. The field strength $B_{ip}$ is
indicated
exemplarily. In-plane magnetoresistance $M\hspace{-0.5mm}R_{ip}$ versus
$B_{ip}$ for field
direction parallel ($\Theta = 0^{\circ}$) and perpendicular
($\Theta = 90^{\circ}$) to current flow direction for (c) bridge A and (d)
bridge E. (e) Anisotropic magnetoconductance $AM\hspace{-0.5mm}C$
versus $\Theta$, the angle
between current $I$ and $B_{ip}$ for the $[1\bar{1}0]$ and $[001]$ direction
for
$B = 14$ T and $T=2K$.
(f) The amplitude of the magnetoconductance oscillations $AMC_{max}$
versus $B_{ip}$ for the $[1\bar{1}0]$ and the $[001]$ direction. The minima
of the
magnetoconductance oscillations were found always at
$\Theta = 0^{\circ}, 180^{\circ}$, i.e.,
$B_{ip}$ parallel to the current direction.}
\end{figure}

Measurements of the $M\hspace{-0.5mm}R$ and $AM\hspace{-0.5mm}R$ with magnetic 
field direction normal or parallel
 to the interface, respectively, are often used to characterize SOI in
low-dimensional electron systems. In STO-based 2DES, the Rashba-type SOI
usually leads to a weak antilocalization (WAL) of the charge carrier transport
at low $T$ \cite{2}, resulting in a logarithmic $T$-dependence of
$R_s$ \cite{23}.
However, the quantum coherence can be destroyed by applying moderate magnetic
fields leading to a distinct positive $M\hspace{-0.5mm}R$ \cite{24}.\\
For $T\ge 50 $K the $M\hspace{-0.5mm}R$ of $(110)$ AO/STO is rather small, 
less than 2\%, and
displays no distinct anisotropy with respect to the crystallographic direction.
For $T < 50$K, $M\hspace{-0.5mm}R$ starts to increase with respect to 
amplitude and anisotropy.
 In Fig. \ref{fig3}, $M\hspace{-0.5mm}R$ is shown for the microbridges A$-$E, 
for $T = 10 $K and $2$K.
For $T = 10 $K, $M\hspace{-0.5mm}R$ is positive and amounts to about 10\%.
The $B$-dependence of
$M\hspace{-0.5mm}R$ indicates orbital motion of free carriers due to the 
Lorentz force, i.e.,
classical Lorentz scattering (LS) as the dominant scattering mechanism, where
$M\hspace{-0.5mm}R$ is well described by the Kohler form:
$M\hspace{-0.5mm}R \propto (1/R_0) \times 
\left(\frac{B}{w}\right)^2/(1+\left(\frac{B}{w}\right)^2)$, 
with the zero-field resistance $R_0$\cite{25}. Fits to
the Kohler form are shown by solid lines in Fig. \ref{fig3} (a).
$M\hspace{-0.5mm}R$ displays clear anisotropic behavior with respect to the 
microbridges, showing a systematic increase from A to E. This is very likely 
related to the decrease of the zero-field resistance from A to E, see 
Fig. 2 (a). \\
For $T = 2 $K, an additional contribution to the positive $M\hspace{-0.5mm}R$ 
appears. However, significant changes to $M\hspace{-0.5mm}R$ are restricted 
to the low field region, $B < 8 $T.
As mentioned above, in 2DES charge transport in the diffusive regime is well
described by the 2D WAL theory\cite{23}. The quantum corrections to the 
conductivity arise from the interference of 
electron waves scattered along closed paths in opposite directions. Phase 
coherence is destroyed if the applied magnetic field which results in a phase 
shift between the corresponding amplitudes
exceeds a critical value. An estimation for the field limit 
$B^* = \hbar/(2el_m^2)$ can be deduced from the electron mean free path 
$l_m=\frac{\hbar}{e}\sqrt{2\pi n_s}\mu$
of the sample. For our sample we obtain $l_m = 12$nm which results in a field 
limit of about 2 T. \\
Zeeman corrections to the WAL are taken into account by the Maekawa and 
Fukuyama (MF)
theory \cite{24}, which is usually used to describe the $B$-dependence of the
$M\hspace{-0.5mm}R$ in LAO/STO and AO/STO \cite{2,20}. The parameters of the 
MF-expression are
the inelastic field $B_i$, the spin-orbit field $B_{so}$, and the electron
g-factor which enters into the Zeeman corrections.\\
For $B \le 2$T, $M\hspace{-0.5mm}R$ at $2$K is well described by LS and WAL.
Fits to the data, using the MF-based expression given in \cite{2} in
combination with a Kohler term, are shown in Fig. \ref{fig3} (b) and (c)
by solid lines.\\
The WAL fitting results in parameters $B_i \approx 180$mT and 
$B_{so} \approx 0.6$T. Within the experimental resolution and the limited field range WAL effect appears to be nearly the same for all the microbridges. 
Zeeman corrections to 
$M\hspace{-0.5mm}R$ have been found to play only a minor role for the applied
magnetic fields. The magnitude of $B_i$ and $B_{so}$ are well comparable to 
those found in $(001)$ AO/STO and LAO/STO,
where Rashba-type SOI has been identified as the dominant source of spin orbit
coupling. \\
In comparison to WAL, contributions from LS to $M\hspace{-0.5mm}R$ at 
$2$K are rather small for 
$B < 2$T. However, for $B > 8$T, where WAL can usually be neglected, 
LS dominates $M\hspace{-0.5mm}R$ again. 
Interestingly, in comparison to the anisotropy of $M\hspace{-0.5mm}R$ with 
respect to the 
microbridges for $B > 8$T and at $T = 10$K, the anisotropy of 
$M\hspace{-0.5mm}R$ at 2 K, is 
slightly decreased. In contrast, the anisotropy of $R_0$, $\mu$, and $n_s$ 
with respect 
to the current direction, are well comparable for $T = 10$K and 2K, or even 
slightly larger at 2K and likely do not explain that behavior.
It might be suggested, that Rashba-type SOI not only influences 
$M\hspace{-0.5mm}R$ by 
WAL at low magnetic fields but also at higher fields, where WAL 
should be absent.\\
Anisotropic Rashba splitting was indeed observed by angle resolved
photoemission (ARPES) experiments on $(110)$ STO surfaces \cite{11} and 
discussed for (110) LAO/STO
heterostructures \cite{8,16}.
The influence of SOI and Rashba effect on magnetotransport can be studied 
more specific, 
if the magnetic field is applied parallel to the interface where changes of 
the $M\hspace{-0.5mm}R$ by WAL are negligible. \\
Applying the magnetic field parallel to the interface at an angle $\phi$ with
respect to the $[001]$ direction, $B_{ip}(\phi)$, results in a strong
field-induced directional anisotropy of the resistance $R_s(B,\phi)$, i.e.,
an $AM\hspace{-0.5mm}R$. Fig. \ref{fig4} (a) and (b) document 
$R_s(B_{ip},\phi)$ versus $\phi$ for
different $B_{ip}$ at $T = 2$K  for bridge A and E, respectively.
$R_s(B_{ip},\phi)$ shows a sinusoidal oscillating two-fold anisotropic behavior
with maxima at $\phi \approx 90^{\circ}/270^{\circ}$ and $0^{\circ}/180^{\circ}$
for A and E, respectively. Obviously, the anisotropy depends much stronger on
the angle between the direction of the bridge, i.e., the direction of
current $I$ and $B_{ip}$ than on the crystallographic direction. Maxima of
$R_s(B_{ip},\phi)$ are always observed for $B_{ip}$ parallel to the microbridge,
i.e., current flow direction. \\
Similar anisotropic behavior was also found in $(001)$ LAO/STO and AO/STO.
In the framework of the Drude-Boltzmann theory it was shown, that a
Rashba-type SOI in $(001)$ LAO/STO induces a two-fold non-crystalline
anisotropy in the magnetoconductance \cite{6}, i.e.,
$\Delta\sigma = [\sigma(B_{ip},\Theta)-\sigma(B_{ip},0) ] \propto
sin^2(\Theta)$, where the amplitude of the oscillations should scale for
moderate field strength with the square of the spin-orbit energy, i.e.,
$\Delta\sigma(\Theta=90^{\circ})/\sigma_0 \propto \Delta_{so}^2$, where
$\sigma_0 = \sigma(B_{ip},0)$ and $\Theta$ the angle between $I$ and $B_{ip}$.
Therefore, it is very likely, that the observed anisotropy of
$R_s(B_{ip},\phi)$ is caused by Rashba-type SOI alike. \\
The amplitude of the oscillations of $R_s(B_{ip},\phi)$ increases with
increasing magnetic field reaching an $AM\hspace{-0.5mm}R$ of about 1\% for A, i.e., along
the $[1\bar{1}0]$ direction and 1.4\% for E, parallel to the $[001]$ direction
for $B_{ip} = 14 $T. The different amplitudes likely indicate an anisotropic Rashba-type SOI.
Note that the $AM\hspace{-0.5mm}R$ is about one order of magnitude smaller 
as compared to the $M\hspace{-0.5mm}R$.
Interestingly, $R_s$ first increases with increasing $B_{ip}$ up to about $5$T
and then decreases again.  This behavior is documented in more detail in
Fig. \ref{fig4} (c) and (d) where the in-plane magnetoresistance
$M\hspace{-0.5mm}R_{ip} = [R_s(B_{ip},\Theta) - R_s(0,\Theta)]
/  R_s(0,\Theta)$ is plotted for
A and E versus $B_{ip}$ for field direction parallel ($\Theta = 0^{\circ}$) and
perpendicular ($\Theta = 90^{\circ}$) to the current $I$ direction. \\
The $M\hspace{-0.5mm}R_{ip}$ for $\Theta = 0^{\circ}$ is only slightly
larger compared to $\Theta = 90^{\circ}$. With increasing $B_{ip}$, the
$M\hspace{-0.5mm}R_{ip}$ first
increases, displaying a maximum positive magnetoresistance around $5$T. Then,
the $M\hspace{-0.5mm}R_{ip}$ decreases and even becomes negative for $B_{ip}$ 
above about $10$T. The negative $M\hspace{-0.5mm}R_{ip}$ at large
fields possibly results from spin-polarized bands due to Zeeman effect,
leading to a suppression of interband scattering with increasing
$B$ \cite{27}.\\
Fig. \ref{fig4} (e) shows the anisotropic magnetoconductance
$AMC= [\sigma(B_{ip},\Theta)-\sigma(B_{ip},0)]/\sigma(B_{ip},0)$ versus $\Theta$
for $I \parallel [1\bar{1}0]$ and $I \parallel [001]$ at $B_{ip} = 14 $T and
$T = 2 $K. The maxima of the magnetoconductance oscillations $AMC_{max}$ always
appear at $\Theta = 90^{\circ}$ and $270^{\circ}$, i.e., $B_{ip}$ perpendicular
to the current direction. For the $[001]$ direction $AMC_{max}$  amounts to
about 1.3\% and is distinctly larger compared to that of the $[1\bar{1}0]$
direction ($\approx 0.8\%$). 
The field dependence of the amplitude $AMC_{max}$ for the two orthogonal
directions is shown in Fig. \ref{fig4} (f). Measurable magnetoconductance
appears
for $B_{ip} > 3$T and increases with field to 1.2\% and 0.9\% at $14$T for
the $[001]$ and $[1\bar{1}0]$ direction, respectively. 
Rashba effect seems to increase with increasing $B_{ip}$ and to be anisotropic with respect to crystallographic direction.\\
In contrast to the non-crystalline $AM\hspace{-0.5mm}R$ of $(110)$ AO/STO
shown here, a giant crystalline $AM\hspace{-0.5mm}R$ was reported for $(110)$
LAO/STO - displaying comparable $n_s$ and $\mu$ - with resistance maxima for
$B_{ip}$ along the $[1\bar{1}0]$ direction, independent of current
direction \cite{8}.  \\
\subsection{Theoretical modeling of the electronic band structure and
magnetotransport}
In order to obtain a better understanding of the measured electronic transport,
especially the $AM\hspace{-0.5mm}R$ behavior, we carried out
tight-binding calculations to model the electronic subbandstructure of
$(110)$ AO/STO. Details to the linear combination of atomic orbitals (LCAO)
calculations are given in the Appendix. The calculation yields the
energy bands $E_{\nu,{\bf k}}$ where $\nu$ is the band index and
${\bf k}$ the wave vector in the rectangular Brillouin zone.\\
\begin{figure}
\includegraphics[width=0.7\columnwidth]{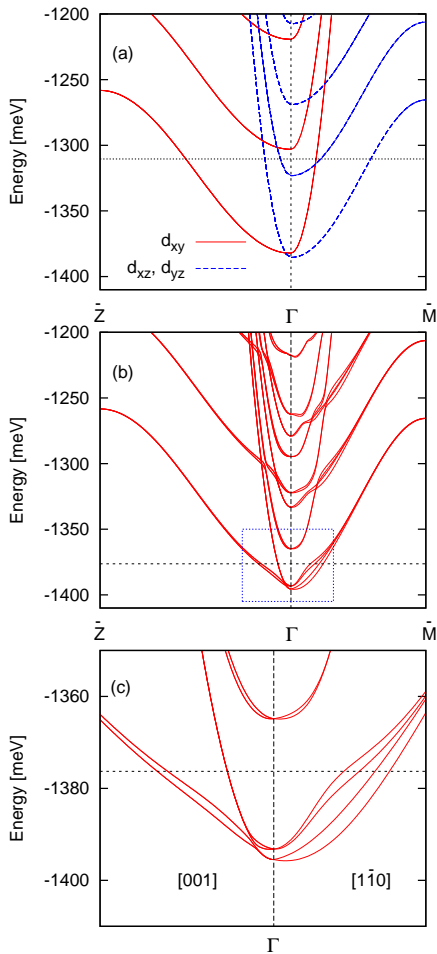}
\caption{\label{fig5}  LCAO Band structure for the $(110)$ AO/STO
interface. (a) Band structure without spin-orbit
coupling and symmetry-breaking electric field.
All bands have two-fold spin degeneracy, the horizonzal line
is the Fermi energy for $n_e=0.4$/unit-cell. (b) Band
structure including spin-orbit
coupling and symmetry-breaking electric field, but $B=0$.
(c) Closeup of the box indicated in (b). In (b) and (c) the horizontal dashed
line is the Fermi energy for $n_e=0.05$/unit-cell.}
\end{figure}
Figure \ref{fig5} shows the band structure obtained in this way.
The topmost panel shows the band structure
in the absence of spin orbit coupling and symmetry breaking electric field,
which roughly agrees with the band structure obtained by Wang {\em et al.}
from a fit to their ARPES data\cite{11} (the reason for the deviation
is our modification of the nearest neighbor hopping $t$ for bonds
in $[001]$-direction, see the discussion in the Appendix).
Since there is no mixing between the three $t_{2g}$ orbitals
the bands can be classified according to the type of $d$-orbital
from which they are composed, whereby the bands derived from the
$d_{xz}$ and $d_{yz}$ orbitals are degenerate.
The dashed horizontal line gives the Fermi energy
for an electron density of $0.40$/unit-cell or
$1.8\times 10^{14} \;$ cm$^{-2}$. This is considerably
higher than the electron densities studied here but
corresponds roughly to the experiments by Wang {\em et al.}\cite{11}.
In the Figure one can identify
the various subbands generated by the confinement of the electrons
perpendicular to the interface. This hierarchy of subbands in fact extends to
considerably higher energies than shown in the Figure.\\
The two lower panels show the band structure for finite spin-orbit coupling
and symmetry breaking electric field, but $B=0$.
One can recognize the two different manifestations of the Rashba effect
discussed already by Zhong {\em et al.} \cite{Zhong}: the splitting of bands
near $\Gamma$ which can be either $\propto |{\bf k}|$ or $\propto |{\bf k}|^3$
(see below) and the opening of gaps. The formation of gaps
is particularly obvious at $\Gamma$ where
the lowest $d_{xy}$-derived band along $\Gamma-\bar{M}$ combines
with one of the $d_{xz}/d_{yz}$-derived bands along $\Gamma-\bar{Z}$
to form a mixed band whose minimum is shifted upward
by $\approx 20 $ meV.
The  dashed horizontal line in the lower two panels gives $E_F$
for an electron density of $n_e=0.05$/unit-cell or $2.3\times 10^{13}\;$cm$^{-2}$
which is roughly appropriate for our experiment. We have verified that varying
the density in the range $0.04$/unit-cell $\le n_e \le 0.07$/unit-cell
does not have a significant influence on the magnetoresistance
discussed below. From now
on the labeling of bands is according to their energy i.e.
the lowest band is labeled 1 and so on.
\begin{figure}
\includegraphics[width=0.8\columnwidth]{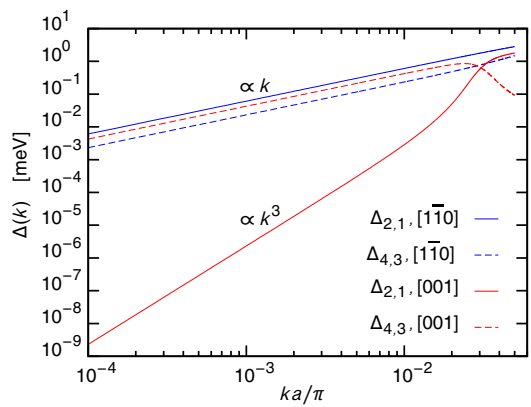}
\caption{\label{fig6}  Rashba-induced band splittings
$\Delta_{\nu,\nu'}({\bf k})$ near $\Gamma$ along the two symmetry lines
of the Brillouin zone.}
\end{figure}
Figure \ref{fig6} shows the differences
$\Delta_{\nu,\nu'}({\bf k})= E_{\nu,{\bf k}}-E_{\nu',{\bf k}}$
and demonstrates
the power-law behaviour of the Rashba-induced band splitting
for small $|{\bf k}|$. Thereby the splitting along $[1\bar{1}0]$
is linear, i.e. $\Delta_{\nu,\nu'}({\bf k})=C_{\nu,\nu'}\frac{ka}{\pi}$
with $C_{2,1}=60.8$ meV and $C_{4,3}=23.4$ meV, whereas
along $[001]$ the splitting between
bands $3$ and $4$ still has this form with $C_{4,3}=42.4$ meV
whereas the splitting between bands $1$ and $2$ now is cubic,
$\Delta_{2,1}(k)=2720$meV$\left(\frac{ka}{\pi}\right)^3$.
This highlights the anisotropy of the Rashba-effect at the 
$(110)$ AO/STO interface.\\
\begin{figure}
\includegraphics[width=\columnwidth]{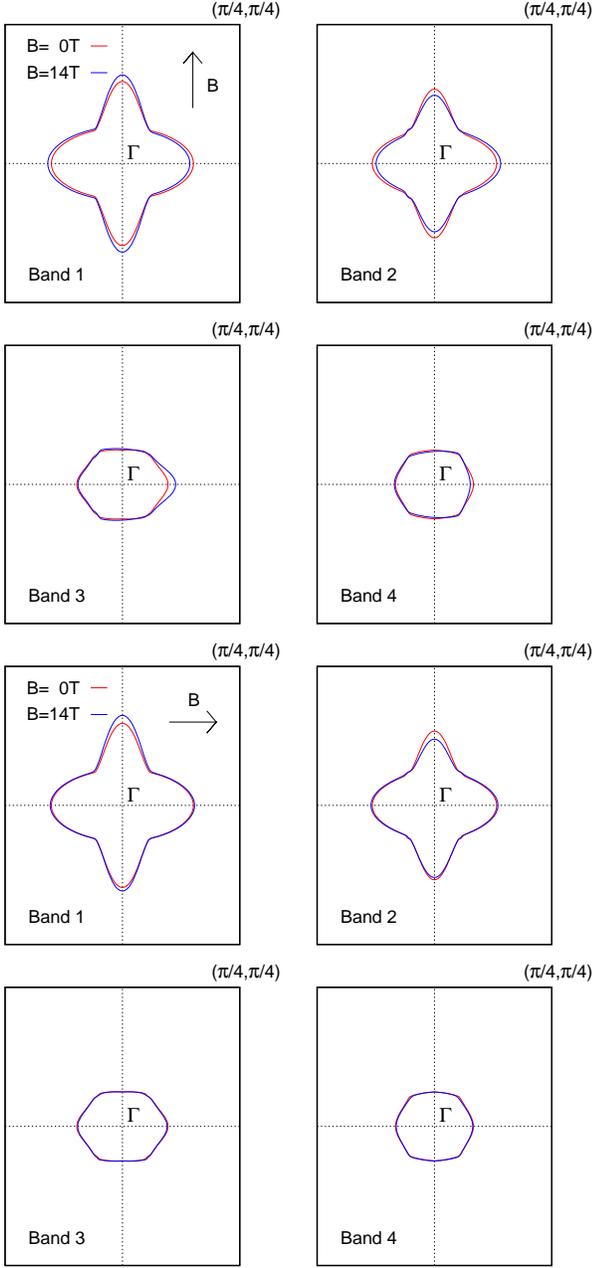}
\caption{\label{fig7}  The four Fermi sheets
for $n_e=0.05/$unit-cell in zero magnetic field and a magnetic field of
$14$ T.
The field direction is $[001]$ for the four topmost panels
and $[1\bar{1}0]$ for the four bottom panels (see arrows).}
\end{figure}
Figure \ref{fig7} compares the Fermi surfaces for $n_e=0.05$/unit
cell in zero magnetic field and a field of $14$ Tesla.
All panels show the square
$[-\frac{\pi}{4}:\frac{\pi}{4}]\otimes[-\frac{\pi}{4}:\frac{\pi}{4}]$,
the field direction is along $[001]$ (along $[1\bar{1}0]$)
in the top four (bottom four) panels.
In the absence of SOC and electric field the Fermi surface
would consist of two elliptical sheets centered at $\Gamma$,
each of them two-fold (spin-)degenerate. The
ellipse derived from the $d_{xz}/d_{yz}$ orbitals is elongated
along the $[1\bar{1}0]$ (or $\Gamma-\bar{M}$) direction, whereas the
ellipse derived from the $d_{xy}$ orbitals is elongated
along the $[001]$ (or $\Gamma-\bar{Z}$) direction. The Rashba effect
splits and mixes these bands and creates the more
complicated $4$-sheet Fermi surface in Figure \ref{fig7}.\\
Switching on the magnetic field results in an area change of
the various Fermi surface sheets as well as
a displacement perpendicular to the field direction whereby
pairs of bands are shifted in opposite direction, namely bands $1$ and $2$
and bands $3$ and $4$. This displacement is considerably
more pronounced for the magnetic field in $[001]$-direction
and barely visible for magnetic field in $[1\bar{1}0]$-direction.
Qualitatively this behaviour can be derived from
the simplified single-band model\cite{Raimondi}:
\begin{eqnarray}
H=\frac{p^2}{2m} + \alpha \;{\bm \tau}\cdot ({\bm p}\times{\bm e}_z) -
\omega_s {\bm \tau}\cdot{\bm B}.
\label{raiham}
\end{eqnarray}
Here $\alpha$ is the strength of the Rashba coupling,
$\omega_s=\mu_BB$ and ${\bm \tau}$ the vector
of Pauli matrices.
The magnetic field ${\bm B}=B {\bm e}_B$ is in the $(x,y)$-plane
and it is assumed
that $p_F^2/2m \gg \alpha p_F, \omega_s$ where $p_F$ is the Fermi momentum.
The eigenvalues are
\begin{eqnarray*}
E_{\bm p}^{(\pm)}&=&\frac{p^2}{2m} \pm
|\; \alpha {\bm p} + \omega_s {\bm e}_\perp |\nonumber \\
&\approx& \left\{ \begin{array}{l c}
\frac{p^2}{2m} \pm \omega_s \pm\;\;
\alpha {\bm p}\cdot {\bm e}_\perp,& \alpha p_F \ll \omega_s\\
&\\
\frac{p^2}{2m} \pm \alpha p  \pm \frac{\omega_s}{p}
 {\bm p}\cdot {\bm e}_\perp,& \omega_s\ll \alpha p_F
\end{array} \right.
\end{eqnarray*}
with ${\bm e}_\perp={\bm e}_B\times{\bm e}_z$.
The Fermi momenta for the two sheets can be parameterized by
the angle $\varphi \in [0,2\pi]$:
\begin{eqnarray*}
{\bm p}_F(\varphi) &=&
\left\{ \begin{array}{l c}
\pm m\alpha\; {\bm e}_\perp
+ \left(\;p_F\pm \frac{m\omega_s}{p_F}\;\right) {\bm e}_p,
& \alpha p_F \ll \omega_s\nonumber \\
&\nonumber \\
\pm \frac{m\omega_s}{p_F}\;{\bm e}_\perp
+ \left(\;p_F\pm m\alpha\;\right) {\bm e}_p,& \omega_s\ll \alpha p_F
\end{array} \right.
\end{eqnarray*}
where $p_F=\sqrt{2m E_F}$ and
${\bm e}_p=(\cos(\varphi),\sin(\varphi))$.
In both limiting cases these are two circular sheets with slightly different
radii, displaced in the direction perpendicular to the magnetic field.
More precisely, when looking along ${\bm B}$, for $\alpha > 0$ the larger
(smaller) circle is displaced to the right (left).
For $\alpha < 0$, on the other hand, the
larger (smaller) circle is displaced to the left (right).
Figure \ref{fig7} shows that for ${\bm B}\parallel [001]$
the bands $1$ and $2$ as well as the bands $3$ and $4$ form two such pairs of
Fermi surface sheets which are displaced in opposite direction. Thereby the
direction of displacement indicates that the sheets $1$ and $2$
appear to have an effective $\alpha < 0$ whereas the two inner
sheets $3$ and $4$ have $\alpha > 0$.
As already mentioned the displacement is much smaller
for ${\bm B}\parallel[1\bar{1}0]$ than for ${\bm B}\parallel[001]$
which again shows the pronounced anisotropy of the Rashba-effect in
the more realistic LCAO-Hamiltonian.\\
Using the energy bands $E_{\nu,{\bf k}}$ we calculated the
$2\times 2$ conductivity tensor using the semiclassical
expression
\begin{eqnarray}
\sigma_{\alpha\beta} &=&e^2\;\sum_\nu\;\tau_\nu\;I_{\alpha,\beta}^{(\nu)}\nonumber \\
I_{\alpha,\beta}^{(\nu)} &=&\frac{1}{4\pi^2}\;
\int\;d{\bm k}\;\delta(E_{\nu,{\bf k}} - E_F)\;v_{\nu,\alpha} v_{\nu,\beta}\nonumber \\
&=& \frac{1}{4\pi^2}\;\int_0^{2\pi}d\varphi\;k_{F,\nu}(\varphi)\;
\frac{v_{\nu,\alpha}(\varphi) v_{\nu,\beta}(\varphi) }
{\nabla_{\bm k} E_{\nu}(\varphi)\cdot{\bf e}_{\bf k} }.
\label{semiclassical}
\end{eqnarray}
Here $\alpha,\beta\in\{x,y\}$,
${\bf k}_{F,\nu}(\varphi)$ is the Fermi momentum of the $\nu^{th}$ sheet along the
direction ${\bf e}_{\bf k}=(\cos(\varphi),\sin(\varphi))$, and
${\bf v}_\nu(\varphi)$ is the velocity
$\hbar^{-1}{\bf \nabla}_{\bm k} E_{\nu{\bm k}}$
evaluated at ${\bf k}_{F,\nu}(\varphi)$.
Moreover, $\tau_\nu$ denotes the lifetime
of the electrons in band $\nu$ which
we assume independent of $\varphi$ for simplicity.\\
Figure \ref{fig8} then shows the variation of the `band resolved'
conductivities
with the angle $\phi$ between magnetic field and $[001]$-axis.
\begin{figure}
\includegraphics[width=\columnwidth]{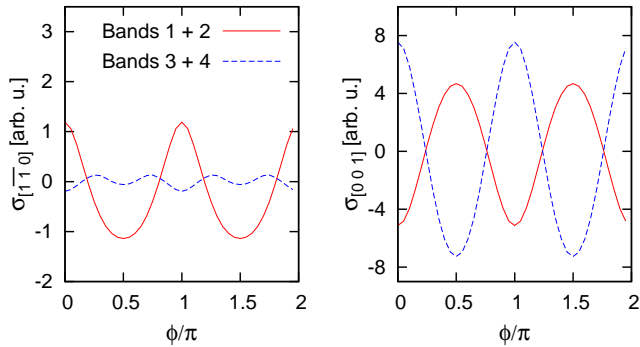}
\caption{\label{fig8}
Variation of $\sigma_{[1\bar{1}0]}$ (left) and $\sigma_{[001]}$ (right)
with the angle  $\phi$ between ${\bf B}$ and
the $[001]$ direction. Thereby $B=14T$.
A $\phi$-independent constant has been subtracted
to make the variations visible, for the labeling of the Fermi surface
sheets see Figure \ref{fig7}.}
\end{figure}
More precisely the figure shows the contributions
of different bands $\nu$ in (\ref{semiclassical})
to the two diagonal elements
$\sigma_{[1\bar{1}0]}$ and $\sigma_{[001]}$ of $\sigma$.
Thereby these contributions are actually summed
over pairs of bands as suggested by Figure \ref{fig7},
which shows that the two sheets belonging to
one pair have similar Fermi surface geometry and shift in opposite
direction in a magnetic field. The variation of the conductivity
with field direction has the form
\begin{eqnarray}
\sigma &\approx& A_0 + A_2\cos(2\Theta) + A_4\cos(4\Theta),
\label{raisimp}
\end{eqnarray}
where $\Theta$ again is the angle between magnetic field and current
direction.
For bands $1$ and $2$ the constant $A_2$ is negative
and substantially larger than $A_4$ so that
the conductivity is minimal for ${\bm j}\parallel {\bm B}$.
This behaviour can be reproduced qualitatively already in the framework
of the generic model (\ref{raiham}).
In the limit $\alpha p_F \ll E_F, \omega_s$ evaluation of
(\ref{semiclassical}) yields
\begin{eqnarray}
\sigma&=&e^2\tau\;\pi^{-1}\;\left[\; E_F - m\alpha^2\;\sin^2(\Theta)\;\right].
\label{raicond}
\end{eqnarray}
Numerical evaluation shows that this result is quite general, i.e. $\sigma$
has the form (\ref{raisimp}) with $A_2<0$ and $A_4=0$ for any $\alpha$ or
$\omega_s$. Figure \ref{fig9} shows the numerical values of $A_2/A_0$ versus
$\omega_s$. Nonvanishing magnetoresistance occurs only
above a threshold value $\omega_s^{(min)}$ which depends on $\alpha$.
This was found previously by Raimondi {\em et al.}\cite{Raimondi}
although these authors did not consider the detailed
variation with field direction.\\
\begin{figure}
\includegraphics[width=0.7\columnwidth]{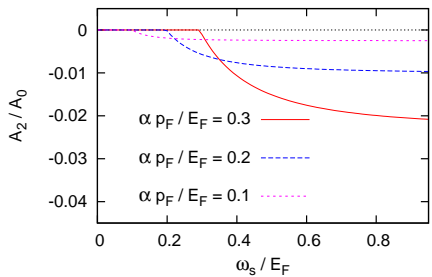}
\caption{\label{fig9}  Ratio of Fourier coefficients
$A_2/A_0$ in (\ref{raisimp}) calculated numerically for the simplified model
(\ref{raiham}).}
\end{figure}
The behaviour of the contribution from
bands $3$ and $4$ differs strongly from the prediction of the simple model.
First, the variation of $\sigma_{[1\bar{1}0]}$ has a substantial admixture
of the higher angular harmonic $\cos(4\Theta)$. Second, while the variation of
$\sigma_{[001]}$ does have a predominant $\cos(2\Theta)$ behaviour,
one now has $A_2 > 0$. The deviating behaviour for this pair of bands
is hardly surprising in that Figure \ref{fig7}
shows that the displacement of the Fermi surface is practically
zero for ${\bm B}\parallel [1\bar{1}0]$ but quite strong for
${\bm B}\parallel [001]$ which suggests that for these two bands
the effective Rashba parameter $\alpha$ depends on
the direction of the magnetic field.\\
Despite an extensive search we were unable to find
a set of LCAO parameters such that the
sheet resistivities (obtained by inversion of the $2\times 2$
conductivity matrix (\ref{semiclassical})) obtained with a single,
band independent
relaxation time $\tau$ match the experimental $R_s$ vs. $\phi$
curves in Figure \ref{fig4}. Agreement with experiment could be achieved only
by choosing a band-dependent relaxation time, more precisely
the relaxation time $\tau_{1,2}$ for the bands $1$ and $2$ had to be chosen
larger by roughly a factor $4$ as compared to $\tau_{3,4}$ for bands
$3$ and $4$. The relaxation times obtained by fitting the experimental data
are shown in Figure \ref{fig10}. They have the expected
order of magnitude and their monotonic and smooth variation with magnetic field
is a few percent. Using the Fermi surface averages
of the Fermi velocity of $\approx 5\times 10^4\;\frac{m}{s}$
the mean free paths are
$l_{1,2}\approx 9$nm and $l_{3,4}\approx 2.5$nm.\\
\begin{figure}
\includegraphics[width=0.7\columnwidth]{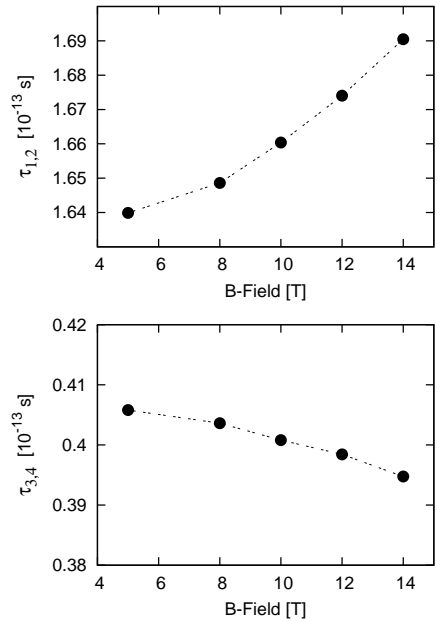}
\caption{\label{fig10}
Variation of the lifetimes $\tau_{1,2}$ and $\tau_{3,4}$
with magnetic field.}
\end{figure}
The resulting $\phi$-dependence of the sheet resistance
is compared to the experimental data
in Figure \ref{fig11}. While the agreement for
current along $[001]$ is good, there is some discrepancy
for current along $[1\bar{1}0]$ in that
the experimental curves have wide minima and sharp maxima,
whereas this is opposite for the calculated curves.
\begin{figure}
\includegraphics[width=\columnwidth]{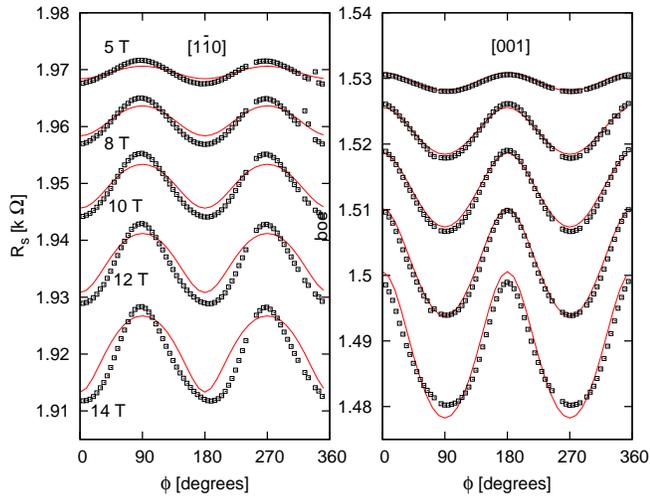}
\caption{\label{fig11}  Calculated
sheet resistance (lines) versus angle $\phi$ between
magnetic field and $[001]$ direction compared to the
experimental data (squares), c.f. Figure \ref{fig4}.}
\end{figure}
This may indicate the limitations of the
quasiclassical Boltzmann approach as already discussed in Ref. \cite{27}
where a full solution of the Boltzmann equation was necessary to reproduce
the experimental data. For completeness
we note that a band dependent relaxation time
has been observed experimentally in materials such
as MgB$_2$\cite{Yelland} and some iron-pnictide
superconductors\cite{Terashima2,Terashima}.\\
Summarizing this section we may say that while a detailed
fit of the experimental data in the framework of the relaxation time
approximation to semiclassical Boltzmann theory is not entirely successful,
the overall behavior observed in experiment - a variation of the
conductivity with magnetic field direction
predominantly according to $\sigma=A_0 + A_2\;\cos(2\Theta)$ i.e. a
noncrystalline anisotropy, is quite generic and can be reproduced 
qualitatively already in the simplest model (\ref{raiham}). Interestingly
the considerably more complicated and 
anisotropic band structure does not change this significantly.
On the other hand, the results show that the
Rashba effect in the (110) surface is strongly anisotropic so that
some deviations from this simple behaviour are not surprising.
The calculation moreover shows that at least within the framework of the
relaxation time approximation the lifetime for electrons in the
two inner Fermi surface sheets must be chosen shorter.
\section{Summary}
Anisotropic electronic transport of the 2DES in $(110)$ AO/STO was
characterized by temperature and magnetic-field dependent 4-point resistance
measurements along different crystallographic directions. 
Anisotropic
behavior of $R_s$ is evident over the complete measured $T$-range
($2$K$\le T \le 300$K)
with lowest sheet resistance and largest electron mobility along the $[001]$ 
direction. The anisotropy
of $\mu$ is mainly responsible for the anisotropic behavior of the normal
magnetotransport $M\hspace{-0.5mm}R$ for $30$K$ > T > 5 $K, where lorentz 
scattering dominates magnetotransport.
At $2$K and $B < 2$T,  $M\hspace{-0.5mm}R$ is dominated by weak 
antilocalization. The spin
orbit field deduced from WAL is well comparable to that found in (001) 
AO/STO and LAO/STO and seems to depend not on specific crystallographic 
direction.\\
Tight-binding
calculations were carried out to model the electronic subbandstructure,
confirming the anisotropy of $\mu$. Despite the high anisotropy
of the Fermi surfaces, the $AM\hspace{-0.5mm}R$ shows a 
non-crystalline behavior with
resistance maxima for in-plane magnetic field parallel to current direction.
Semi-classical Boltzmann theory was used to calculate conductivity and 
$AM\hspace{-0.5mm}R$
confirming the rather unexpected experimental result of a non-crystalline
$AM\hspace{-0.5mm}R$, despite strong anisotropic Fermi surface sheets
and Rashba coupling which however lead to
a strong sensitivity of the $AM\hspace{-0.5mm}R$
behavior of $(110)$ AO/STO on $E_F$ as already observed for $(001)$ LAO/STO.
On the other side, electronic subband-engineering by, e. g., epitaxial strain,
may also provide possibilities to tune $AM\hspace{-0.5mm}R$ behavior which 
might be interesting with respect to spintronics.\\

ACKNOWLEDGEMENTS\\
Part of this paper was supported by the Deutsche Forschungsgemeinschaft (DFG)
Grant No. FU 457/2-1. We are grateful to R. Thelen and the Karlsruhe Nano
Micro Facility (KNMF) for technical support. We also acknowledge D. Gerthsen
and M. Meffert from the laboratory for electron microscopy (LEM) for
transmission electron microscopy analysis of our samples.
\section{Appendix: LCAO calculation}
We describe bulk SrTiO$_3$ as a simple cubic lattice of Ti atoms with lattice
constant unity
at the positions ${\bm R}_i$ and retain only the three $t_{2g}$ orbitals
of each Ti atom.
The Hamiltonian is most easily formulated in a coordinate system
with axes parallel to Ti-Ti bonds which we call the bulk coordinate
system. In the following
$\alpha,\beta,\gamma \in\{x,y,z\}$ always refer to the
bulk coordinate system, are assumed to be pairwise
unequal, and
${\bm e}_\alpha$ denotes the lattice vector in $\alpha$-direction.
Following Wang {\em et al.}\cite{11}
we use a tight-binding parameterization of the Hamiltonian
with hopping integrals
\begin{eqnarray}
\langle d_{\alpha\beta}({\bm R}_i \pm {\bm e}_\alpha)|H| d_{\alpha\beta}({\bm R}_i)
\rangle &=& t,\nonumber \\
\langle d_{\alpha\beta}({\bm R}_i \pm {\bm e}_\gamma)|H|d_{\alpha\beta}({\bm R}_i)
\rangle &=& t_1,\nonumber\\
\langle d_{\alpha\beta}({\bm R}_i \pm {\bm e}_\alpha\pm {\bm e}_\beta)|H|
d_{\alpha\beta}({\bm R}_i)
\rangle &=& t_2.
\label{lcao}
\end{eqnarray}
Following Zhong {\em et al.}\cite{Zhong}
we model the interface as a hemispace of bulk SrTiO$_3$
with surface perpendicular to the unit vector
${\bm e}_n=\frac{1}{\sqrt{2}}(1,1,0)$ and the origin of the
coordinate system coinciding with some atom on the surface.
Accordingly, only atoms with ${\bm R}_i \cdot {\bm e}_n \ge 0$ are retained.
The electrons are confined to the interface by a wedge-shaped
electrostatic potential which gives an extra energy
$\epsilon_i= eE {\bm R}_i \cdot {\bm e}_n$ with $E> 0$ for
all three $t_{2g}$ orbitals on the Ti atom at ${\bm R}_i$.\\
The unit-cell of the resulting $(1,1,0)$ surface is a rectangle
with edges $\sqrt{2} \parallel [1\bar{1}0]$ and $1\parallel [001]$.
The Brillouin zone has the extension $\sqrt{2}\pi$ in
$[1\bar{1}0]$-direction and $2\pi$ in $[001]$-direction and we define
$\bar{M}=(\pi/\sqrt{2},0)$ and $\bar{Z}=(0,\pi)$. \\
Using the model described so far, Wang {\em et al.}
obtained an excellent fit to their ARPES band structure at
the SrTiO$_3$ $(110)$ surface by using the values
$t=-277$ meV, $t_1=-31$ meV, $t_2=-76$ meV
and $eE=10\;meV/\sqrt{2}$. Using these values, the conductivity
calculated within the Boltzmann equation formalism (as described in the
main text)
shows a rather strong anisotropy,
$\sigma_{[001]}/\sigma_{[1\bar{1}0]}\approx 3.5$,
much larger than the experimental value
$\sigma_{[001]}/\sigma_{[1\bar{1}0]} \approx 1.3$.
The reason is that for the low electron densities of $\approx 0.05$/unit
cell in our experiment the $d_{xy}$-derived band which has small effective
mass along the $[1\bar{1}0]$ direction (see Figure \ref{fig5})
is almost empty. This can be changed, however,
by reducing $t\rightarrow-269.5$ meV for all bonds in $[001]$-direction.
This reduction by $2.5\%$ might be the consequence of a slight distortion
of the lattice in the neighborhood of the interface. The same reduction
of the anisotropy could also be obtained by lowering the energy of the
$d_{xy}$-orbital by $\approx 10$ meV. Both modifications
shift the minimum of the $d_{xy}$-derived band to lower energy
and thus increase its filling.\\
To discuss the magnetoconductance
we extended the model of Wang {\em et al.} by including the Rashba effect -
that means the combined effect of spin-orbit coupling in the Ti 3d shell
and the confining electric field - as well as an external magnetic field.
First, the nonvanishing matrix elements
of the of orbital angular momentum operator
${\bm L}$ within the subspace of the $t_{2g}$ orbitals are
\begin{eqnarray*}
\langle d_{xz}|L_x |d_{xy}\rangle &=& i\hbar,
\end{eqnarray*}
plus two more equations obtained by cyclic permutations of $(x,y,z)$.
Choosing the basis on each Ti atom
as $(d_{xy,\uparrow},d_{xy,\downarrow},d_{xz,\uparrow},d_{xz,\downarrow},
d_{yz,\uparrow},d_{yz,\downarrow})$
one thus finds
\begin{eqnarray*}
L_x&=&\hbar\;\left(\begin{array}{r r r}
 0 &-i & 0\\
 i & 0 & 0\\
 0 & 0 & 0 \\
\end{array}\right)\otimes \tau_0,\\
L_y&=&\hbar\;\left(\begin{array}{r r r}
 0 & 0 & i\\
 0 & 0 & 0\\
-i & 0 & 0 \\
\end{array}\right)\otimes \tau_0,\\
L_z&=&\hbar\;\left(\begin{array}{r r r}
0 &  0 & 0\\
0 &  0 &-i\\
0 &  i & 0 \\
\end{array}\right)\otimes \tau_0,\\
\end{eqnarray*}
where $\tau_0$ is the unit matrix in spin space.
The Hamiltonian for the spin orbit-coupling then is\cite{Zhong}
\begin{eqnarray*}
H_{SO} &=&\lambda_{SO}\;{\bm L}\cdot {\bm S}\\
&=&\frac{\lambda_{SO}\;i\hbar^2}{2}\;\left(\;
\left(\;|xz\rangle\langle xy| - |xy\rangle\langle xz|\;\right)\;\tau_x
+ c.p.\;\right)\\
&=&\frac{\lambda_{SO}\;\hbar^2}{2}\;
\left(\begin{array}{r r r}
0        & -i\tau_x & i\tau_y\\
 i\tau_x &     0    &-i\tau_z\\
-i\tau_y &  i\tau_z & 0
\end{array}\right).
\end{eqnarray*}
Here $c.p.$ denotes two more terms obtained by cyclic
permutation of $x,y,z$ and ${\bm \tau}$ is the vector
of Pauli matrices.
We use $\lambda_{SO}\hbar^2 =20$ meV.
The coupling to an external magnetic field ${\bm B}$ is
\begin{eqnarray*}
H_B&=&\mu_B({\bf L} + g \;{\bf S})\cdot {\bm B}
\end{eqnarray*}
with the Bohr magneton $\mu_B$ and
we use $g=5$\cite{fete}.\\
In addition to the matrix elements (\ref{lcao}) the confining electric field
gives rise to small but nonvanishing hopping elements, which
would vanish due to symmetry in the bulk.
The respective term in the Hamiltonian is
$H_E=|e|{\bf E}_\perp\cdot {\bf r}$ where
${\bf E}_\perp$ is the component of the electric field perpendicular
to the bond. As shown by Zhong {\em et al.}\cite{Zhong} the respective
matrix elements can be written as ($\alpha$, $\beta$ and $\gamma$
refer to the bulk system and are pairwise unequal)
\begin{eqnarray*}
\langle d_{\alpha\beta}({\bm R}_i \pm {\bm e}_\gamma) | H_E |
d_{\beta\gamma}({\bm R}_i)\rangle &=& \pm |e|\;E_\alpha\;V_E,\\
\langle d_{\beta\gamma}({\bm R}_i \pm {\bm e}_\gamma) | H_E |
d_{\alpha\beta}({\bm R}_i)\rangle &=&  \mp|e|\;E_\alpha\;V_E,
\end{eqnarray*}
and we used the value $|e|\;E\;V_E =5$ meV. The sign of
$V_E$ is positive if one really considers only
two $d$-orbitals at the given distance. This might change if one
really considers hopping via the oxygen-ion between the two
Ti ions in the true crystal structure of SrTiO$_3$. 
We have verified, however,that inverting the sign of $V_E$ does not 
change the angular dependence of the magnetoresistance in 
Figure \ref{fig11}, although it does in fact
change the direction of the shift of the Fermi surface sheets in
Figure \ref{fig3}, that means the sign of the effective
$\alpha$. In fact, as can be seen from (\ref{raicond})
the sign of $\alpha$ does not influence the angular variation
of the conductivity.\\
We neglect any matrix elements of the electric field
between orbitals centered on atoms more distant than
nearest neighbors. The interplay between these additional hopping
matrix elements and the spin orbit coupling gives rise to the
Rashba splitting of the bands.
Adding the respective terms to the tight-binding Hamiltonian
we obtain the band structure and its variation with a magnetic field.
We have verified that slight variations of $\lambda_{SO}$, $|e|EV_E$
or $g$ do not lead to qualitative changes of the results reported in the
main text.

\end{document}